\documentclass[twocolumn,trackchanges]{aastex62}

\shortauthors{Zongyin Wu et al.}
\begin{document}
\title{The Observations of Magnetic Reconnection During the Interaction Process of Two Active Region Filaments}

\correspondingauthor{Zhike Xue}
\email{zkxue@ynao.ac.cn}

\author[0000-0003-2045-8994]{Zongyin Wu}
\affiliation{Yunnan Observatories, Chinese Academy of Sciences, Kunming 650216, China}
\affiliation{University of Chinese Academy of Sciences, Beijing 100049, China}

\author{Zhike Xue}
\affiliation{Yunnan Observatories, Chinese Academy of Sciences, Kunming 650216, China}
\affiliation{Yunnan Key Laboratory of Solar Physics and Space Science, Kunming 650216, China}

\author{Xiaoli Yan}
\affiliation{Yunnan Observatories, Chinese Academy of Sciences, Kunming 650216, China}
\affiliation{Yunnan Key Laboratory of Solar Physics and Space Science, Kunming 650216, China}

\author{Jincheng Wang}
\affiliation{Yunnan Observatories, Chinese Academy of Sciences, Kunming 650216, China}
\affiliation{Yunnan Key Laboratory of Solar Physics and Space Science, Kunming 650216, China}

\author{Liheng Yang}
\affiliation{Yunnan Observatories, Chinese Academy of Sciences, Kunming 650216, China}
\affiliation{Yunnan Key Laboratory of Solar Physics and Space Science, Kunming 650216, China}

\author{Zhe Xu}
\affiliation{Yunnan Observatories, Chinese Academy of Sciences, Kunming 650216, China}
\affiliation{Yunnan Key Laboratory of Solar Physics and Space Science, Kunming 650216, China}

\author{Qiaoling Li}
\affiliation{Department of Physics, Yunnan University, Kunming 650091, China}

\author{Yang Peng}
\affiliation{School of Economics and Management, Hebei University of Science and Technology, Shijiazhuang 050018, China}

\author{Liping Yang}
\affiliation{School of Physics, Electrical and Energy Engineering, Chuxiong Normal University, Chuxiong 675000, China}

\author{Yian Zhou}
\affiliation{Yunnan Observatories, Chinese Academy of Sciences, Kunming 650216, China}

\author{Xinsheng Zhang}
\affiliation{Yunnan Observatories, Chinese Academy of Sciences, Kunming 650216, China}
\affiliation{University of Chinese Academy of Sciences, Beijing 100049, China}

\author{Liufan Gong}
\affiliation{Yunnan Observatories, Chinese Academy of Sciences, Kunming 650216, China}
\affiliation{University of Chinese Academy of Sciences, Beijing 100049, China}

\author{Qifan Dong}
\affiliation{Yunnan Observatories, Chinese Academy of Sciences, Kunming 650216, China}
\affiliation{University of Chinese Academy of Sciences, Beijing 100049, China}

\author{Guotang Wu}
\affiliation{Yunnan Observatories, Chinese Academy of Sciences, Kunming 650216, China}
\affiliation{University of Chinese Academy of Sciences, Beijing 100049, China}
\begin{abstract}
We investigate the interaction between two filaments (F1 and F2) and their subsequent magnetic reconnection in active region (AR) NOAA 13296 and AR NOAA 13293 on May 9, 2023, utilizing high spatial and temporal resolution and multi-wavelength observational data from the Solar Dynamics Observatory, the New Vacuum Solar Telescope, and the Chinese H$\alpha$ Solar Explorer. The movement of F1 from the southeast toward the northwest, driven by the motion of the positive magnetic polarity (P1), leads to a collision and reconnection with F2. This reconnection exchanges their footpoints, resulting in the formation of two new filaments (F3 and F4) consistent with "slingshot" type filament interaction. During the interaction, the current sheet moving due to the motion of F1 and the reconnection outflows moving along F3 and F4 were both observed. The current sheet is rarely observed in the slingshot type filament interaction, measuring approximately 2.17 Mm in length and 0.84 Mm in width. After the interaction, the F1 disappears whereas a portion of F2 remains, indicating that the interaction involves partial slingshot reconnection, due to the unequal magnetic flux between the filaments. The residual part of F2 will undergo another magnetic reconnection in the same interaction region with the magnetic loops connecting polarities N1 and P1. The material generated by the reconnection is continuously injected into F4, leading to its final morphology. The findings enhance our understanding of slingshot-type filament interactions, indicating that partial slingshot reconnections between filaments may be more common than full slingshot events.

\end{abstract}


\section{Introduction}\label{sec:intro}
  Solar filaments are relatively cooler and denser plasma structures suspended in the solar corona above the magnetic polarity inversion line (PIL), with their footpoints anchored in regions of opposite magnetic polarity. In the solar disk observations, they appear as dark absorption features in strong spectral lines (e.g., H$\alpha$) and extreme ultraviolet (EUV) wavelengths (e.g., 304 Å, 193 Å). However, they appear as bright features at the solar limb due to the dark background, and are also known as prominences \citep{Engvold...1976SoPh...49..283E,Martin...1998SoPh..182..107M,Parenti...2014LRSP...11....1P}. According to their location, solar filaments can be categorized into three classifications: active region (AR) filaments, intermediate filaments, and quiescent filaments \citep{Engvold...1998ASPC..150...23E,Mackay...2010SSRv..151..333M}. Active region filaments are often located in ARs with sunspot activity, while intermediate filaments are situated at the boundaries of active regions. Quiescent filaments are found in the quiet region and polar crown regions. \cite{Martin...1994ASIC..433..303M} and \cite{Martin...McAllister..1996mpsa.conf..497M} discovered the presence of chirality, or handedness property, in filaments and their corresponding coronal arcades. It is that a filament is referred to as dextral (sinistral) if the axial field of the filament points to the right (left) when observed from the side of the positive polarity. \cite{chenpf2014ApJ...784...50C} proposed an another method for determining filament chirality based on skewness, and found that right-skewed drainage corresponds to sinistral, and left-skewed to dextral. The two types of magnetic topology for filaments that have been proposed are the sheared arcade  \citep{Kippenhahn...Schluter...1957ZA.....43...36K,Antiochos...Dahlburg...1994ApJ...420L..41A,DeVore...Antiochos...2000ApJ...539..954D} and the flux rope \citep{Kuperus...Raadu...1974AandA....31..189K,YanXL...2014ApJ...797...52Y,Chengxin...2014ApJ...789L..35C,Priest...Longcope...2017SoPh..292...25P,Tan...Song...2022MNRAS.516L..12T,YangLP...2023ApJ...943...62Y}. Filament configurations are believed to form through two mechanisms: the surface mechanism (e.g., vortical motions, shearing motions, magnetic field convergence, cancellation and reconnection) \citep{van...Ballegooijen...Martens...1989ApJ...343..971V,DeVore...Antiochos...2000ApJ...539..954D,Martens...Zwaan...2001ApJ...558..872M,Amari...2011ApJ...742L..27A,YanXL...2015ApJS..219...17Y,WangJincheng...2017ApJ...839..128W} and subsurface mechanism \citep{Gibson...2004ApJ...617..600G,Okamoto...2008ApJ...673L.215O,Okamoto...2009ApJ...697..913O,YanXL...2017ApJ...845...18Y}. Some studies report that interactions between filaments may lead to the formation of new filaments \citep{Su...2007SoPh..242...53S,Jiang...2014ApJ...793...14J,Joshi...2014ApJ...795....4J,Luna...2017ApJ...850..143L,Koleva...2022SoPh..297...44K}. The material produced by magnetic reconnection during the interaction is injected into the newly  filaments \citep{Su...2007SoPh..242...53S,Xue...2016NatCo...711837X}. 

  Filament interactions occur when two filaments approach each other and collide \citep{Kumar...2010ApJ...710.1195K}. \cite{Linton...2001ApJ...553..905L} studied the interaction between two isolated, highly twisted flux tubes that was simulated in three dimensions.  Based on their relative twists and axial orientations, four types of flux tube interactions are proposed: bounce, merge, slingshot, and tunnel. Their simulation results showed that the slingshot interaction occurred when two counter-helicity flux ropes collided at a sufficiently large angle. However, in a subsequent study by \cite{Linton...2006JGRA..11112S09L}, it was found that the collision of two lowly twisted flux tubes with identical helicity also resulted in the slingshot interaction. Due to the unequal axial flux of two flux tubes, their collision may only result in a partial slingshot interaction: the flux tube with smaller flux may only reconnect with the outer shell of the flux tube with larger flux and leaving behind a unreconnected core flux suspended between the mutually reconnected flux tubes. Both of the above simulation studies were conducted under convective zone conditions. However, it should be noted that filament interactions primarily take place in the solar corona.

The interaction of filaments has been confirmed by many observations, providing insights into the formation and dynamic evolution of filaments. In the corona, three types of filament interactions have been observed: merge, slingshot, and bounce. Interactions between two filaments of the same chirality have often resulted in the coalescence into a single filament \citep{Schmieder...2004SoPh..223..119S,Jiang...2014ApJ...793...14J,YangBo...2016ApJ...830...16Y,Zheng...Ruisheng...2017ApJ...836..160Z,Luna...2017ApJ...850..143L}. \cite{Schmieder...2004SoPh..223..119S} reported the merging event of two filaments due to magnetic shear proximity, with their numerical simulations and conditions for dark filament merging being further refined by others \citep{DeVore...2005ApJ...629.1122D,Aulanier...2006ApJ...646.1349A}. \cite{van..Ballegooijen..2004ApJ...612..519V} observed that two filaments approached and merged into a single elongated U-shaped filament, with magnetic reconnection playing a significant role during both the formation process and subsequent evolution. In the plasma simulations conducted by \cite{Gekelman...2012ApJ...753..131G}, when two flux ropes collide, a reverse current layer is generated between them, leading to magnetic reconnection. This reconnection process can be identified by locating the QSLs, exhibiting characteristics of locality and intermittency. Furthermore, filament interactions can also occur between a lower filament and an upper one through their convergences,  and result in a flare during their eruption \citep{Zhu...2015ApJ...813...60Z}. However, in the filaments slingshot interaction, the two filaments are not merged; instead, their footpoints are exchanged to form two new filaments that move away from the interaction region. Utilizing the H$\alpha$ data on November 20, 2003, \cite{Kumar...2010ApJ...710.1195K} and \cite{Chandra...2011SoPh..269...83C} observed the interaction, reconnection, and footpoint connectivity changes between two adjacent filaments. \cite{Torok..2011yes...ApJ...728...65T} later simulated the same event under coronal conditions and revealed that slingshot reconnection could occur between two coronal flux ropes. \cite{Filippov...2011ARep...55..541F} provided an example to demonstrate the connectivity changes between filaments with the identical chirality. However, no brightening or flare-like phenomena were observed during the interaction between the filaments. Furthermore, \cite{Jiang...2013ApJ...764...68J} reported a slingshot interaction event caused by two filaments with identical chirality. In the three events mentioned above, the two interacting filaments are of the same chirality. However, \cite{Yang...2017ApJ...838..131Y} presented evidence of interactions resulting from magnetic reconnection between two colliding filaments which had a contact angle of approximately 90°, thereby supporting the slingshot interaction theory proposed by \cite{Linton...2001ApJ...553..905L}. These research results suggest that slingshot interactions can take place between two interacting filaments, regardless of whether they have the identical chirality or non-identical chirality. The bounce interaction, which occurs when two filaments undergo weak reconnection and then rebound, was observed by \cite{YangLP...2023ApJ...943...62Y}, where the two interacting filaments have opposite helicity and the contact angle between them is less than 45°.

Observational evidence of magnetic reconnection between two filaments has been observed in several events, accompanied by the formation of two new filaments \citep{Yang...2017ApJ...838..131Y} and bidirectional jets \citep{Shen...2019ApJ...883..104S,YangBo...2019ApJ...887..220Y,YangLP...2023ApJ...943...62Y}. However, the fine process of the reconnection is rarely reported. In this study, we present a detailed analysis of the interactions and reconnection between two filaments using the high-resolution and multi-wavelength observational data from multiple observational equipments. Section \ref{sec:Observations and methods} shows the observational data and methodologies. The main results are presented in Section \ref{sec:results}, and the conclusions and discussions are given in Section \ref{sec:conclusions and discussions}.

\section{OBSERVATIONS AND METHODS}\label{sec:Observations and methods}

The high-resolution data were primarily obtained by the New Vacuum Solar Telescope \citep[NVST;][]{LiuZhong...2014RAA....14..705L}, the Chinese H$\alpha$ Solar Explorer \citep[CHASE;][]{LiChuan...2022SCPMA..6589602L}, the Solar Dynamics Observatory \citep[SDO;][]{Pesnell...2012SoPh..275....3P}. The NVST is a 1 m ground-based observatory situated in the Fuxian Solar Observatory of Yunnan Observatories, Chinese Academy of Sciences. It is designed to detect small-scale structures and solar activity in both the photosphere and chromosphere \citep{Yan...2020b..2020ScChE..63.1656Y}. The NVST is equipped with multiple observation channels, including H$\alpha$, TiO, and He 10830 Å. In this study, the H$\alpha$ line center image at 05:49:00 UT on May 9, 2023 is used to analyze the position and morphology of the newly formed filaments by the magnetic reconnection. On October 14, 2021, the CHASE satellite was successfully launched into a sun-synchronous orbit from the Taiyuan Satellite Launch Center using the Long March 2D rocket. The H$\alpha$ Imaging Spectrograph (HIS) onboard CHASE conducts full-disk spectroscopic observations in the H$\alpha$ (6559.7–6565.9 Å) and Fe I (6567.8–6570.6 Å) wavelength bands with raster scanning mode. The spectrograph data provides a spectral resolution of 0.024 Å per pixel, a spatial resolution of 0.''52 per pixel, and a temporal resolution of 60 seconds. The image at the H$\alpha$ line center (6562.82 Å) is utilized to demonstrate the position and structure of the filaments before the magnetic reconnection. The line-of-sight (LOS) magnetograms and magnetic vector fields were acquired using the Helioseismic and Magnetic Imager \citep[HMI;][]{Schou...2012SoPh..275..229S} aboard the SDO. By utilizing the 304, 171, and 94 Å images observed by the Atmospheric Imaging Assembly \citep[AIA;][]{Lemen...2012SoPh..275...17L} aboard the SDO, we investigate the complete evolution of the filament interaction as well as the relevant magnetic reconnection features. The all images taken from NVST, CHASE, and SDO are aligned by differentially rotating to the reference time of 02:00 UT on May 9, 2023.

The thermodynamic properties of the current sheet are diagnosed using the Differential Emission Measure (DEM) method \citep{Cheung...2015ApJ...807..143C,Su...2018ApJ...856L..17S}. It employs data from six extreme ultraviolet (EUV) wavelengths (94, 131, 171, 193, 211, and 335 Å) sourced from the SDO/AIA. To improve the signal-to-noise ratio (\( \frac{S}{N} \)), we average image intensities over one minute during the DEM analysis. The total Emission Measure (EM) is expressed as:

\[
EM = \int n_e^2 dl,
\]
where \( n_e \) represents the electron number density, and \( l \) denotes the optical depth along the LOS.

\section{RESULTS}\label{sec:results}
\begin{figure}
\plotone{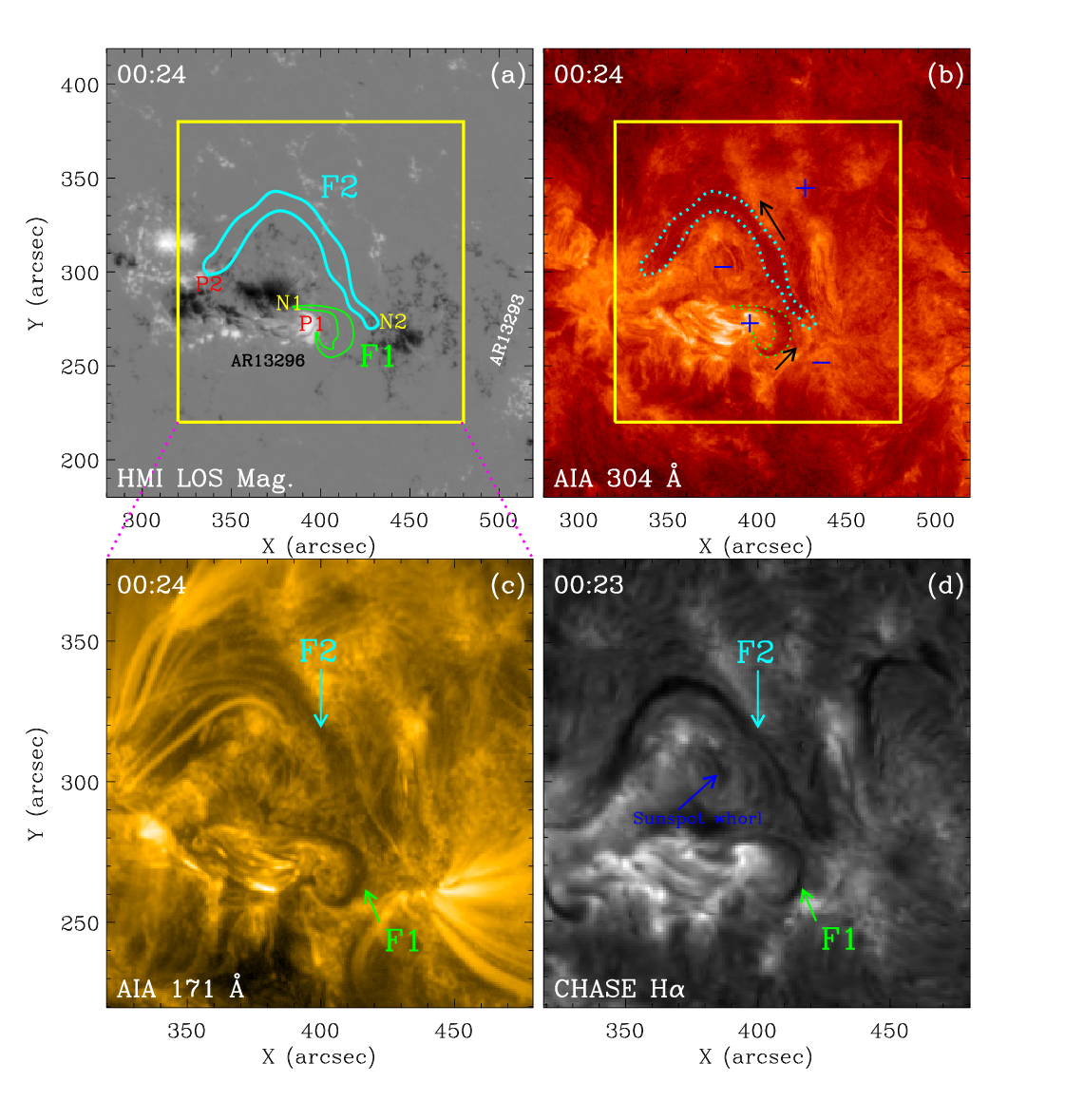}
\setlength{\abovecaptionskip}{-0.7cm}
\caption{(a) SDO/HMI LOS magnetogram, in which positive (negative) magnetic fields are shown in white (black). (b) AIA 304, (c) 171 Å, and (d) CHASE H$\alpha$ image.  In panels (a) and (b), the green and cyan profiles denote the two filaments F1 and F2, respectively. The magnetic polarities corresponding to the footpoints of F1 and F2 are labeled as "N1," "P1," "N2," and "P2," which can be seen before the interaction in panel (a). In panel b, the directions of the axial field of the two filaments are indicated by black arrows. In panel (b), the directions of the axial field of the two filaments are indicated by black arrows.} In panels (c),  and (d), green and cyan arrows indicate the locations of the two filaments, labeled “F1” and “F2,” respectively.
\label{fig:fig1}
\end{figure}

  On May 9, 2023, the interacting filaments were located in the ARs NOAA 13296 (N16W34) and NOAA 13293 (N10W47), situated in the northwestern quadrant of the solar disk. Figure \ref{fig:fig1} illustrates the general appearance of the filaments and the interaction region prior to magnetic reconnection. F1 was situated to the southwest of AR13296, while F2 was a longer filament oriented from northeast to southwest, positioned to the north of F1.  F1's footpoints were anchored in the positive polarity (P1) and the main negative polarity (N1), and F2's ones in  the another positive polarity (P2) and the negative polarity (N2). When the pre-interactive 304 Å outlines of the filament axes are superimposed on the corresponding LOS magnetogram (Figure \ref{fig:fig1}(a)), it is clear that they are aligned along two distinct magnetic PILs. In the AIA 304 Å and 171 Å images (Figures \ref{fig:fig1}(b) and (c)), the contours of F1 and F2 are wider and larger compared to the H$\alpha$ filaments (as evidenced in Figure \ref{fig:fig1}(d)) \citep{Schwartz...2004AA...421..323S}. The axial field angle between the two filaments is approximately $\frac{5}{4} \pi$. In both the EUV images and the H$\alpha$ images, we can see that F2 is thicker than F1. According to the handedness determination rule by \cite{Martin...1994ASIC..433..303M}, a filament is referred to as dextral (sinistral) if the axial field of the filament points to the right (left) when observed from the side of the positive polarity. In Figure 1(b), the direction of the axial field for the filament is marked (as indicated by the black arrows beside the filament), along with the magnetic field polarity in the surrounding regions. For F1, when viewed from the region of positive polarity, its axial field points to the left; thus, it is classified as a sinistral filament. Similarly, for F2, when viewed from the positive polarity region, its axial field also points to the left, making it another sinistral filament. Therefore, the two filaments have the same chirality and neither conforms to the chirality rule of northern hemisphere filaments \citep{Pevtsov...2003ApJ...595..500P}. In Figure 1(d), the sunspot whorl is inclined to rotate clockwise around the sunspot, suggesting that the sunspot may possess positive helicity \citep{Nakagawa...1971SoPh...19...72N,Rust.and.Martin...1994ASPC...68..337R,Chandra...2010SoPh..261..127C}. This supports the conclusion that both F1 and F2 are sinistral filaments.

\begin{figure}
\plotone{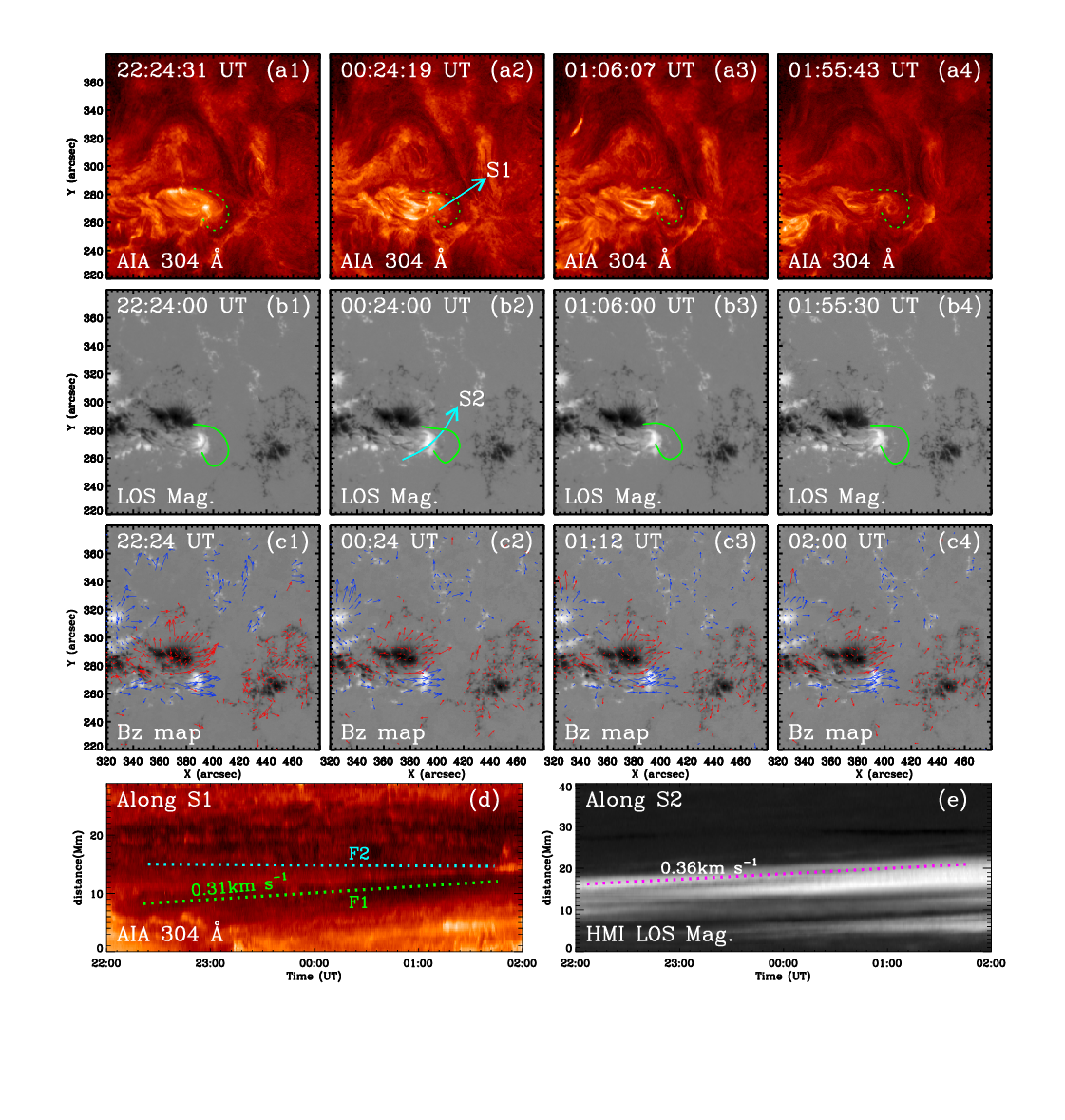}
\setlength{\abovecaptionskip}{-1.5cm}
\caption{The motion of F1 and the evolution of the photospheric magnetic field are illustrated in AIA 304 Å images, LOS magnetograms, and Bz maps. The green lines in panels (a1) to (b4) indicate the position of the filament F1. The contour of F1 is overlaid on the corresponding LOS magnetograms at the same time (panels (b1)-(b4)). Time-slice maps along the cyan lines S1 in panel (a2) and S2 in panel (b2) are presented in panels (d) and (e), respectively. Panels (c1)–(c4) display positive (negative) magnetic fields depicted in white (black). Transverse magnetic fields are represented by small blue (red) arrows superimposed on the positive (negative) polarity regions of the longitudinal magnetic fields. The length and direction of these arrows denote the strength and orientation of the transverse magnetic field. (d) The time-slice map calculated along the cyan line in panel (a2) using AIA 304 Å images. (e) The time-slice map calculated along the cyan line in panel (b2) utilizing the LOS magnetograms. An animation of the SDO/HMI LOS magnetogram, with a field of view comparable to that of panel (a1), is available for the 304 Å images. Its duration spans from 22:00 UT on May 8 to 06:00 UT on May 9.
\label{fig:fig2}}
\end{figure}

To investigate the driving mechanisms of the interaction between the two filaments, we analyzed the 304 Å images (Figure \ref{fig:fig2}a), LOS magnetograms (Figure \ref{fig:fig2}b), and Bz maps (Figure \ref{fig:fig2}c). As shown in Figures \ref{fig:fig2}(a1-a4), F1 gradually approaches F2 along the direction indicated by slice S1 (the cyan line in Figure \ref{fig:fig2}(a2)). To trace the motion of F1, we performed a time-slice map of the 304 Å images along the slice S1, with the results presented in Figure \ref{fig:fig2}(d). The velocity of F1 (marked by the green dashed line in Figure \ref{fig:fig2}(d)) was calculated to be approximately 0.31 km $s^{-1}$. During the approach of F1 to F2, F2 remained relatively stationary (as indicated by the cyan dashed line in Figure \ref{fig:fig2}(d)). In the LOS magnetogram (Figures \ref{fig:fig2}(b1-b4)), P1 is observed moving from the southeast towards the northwest. To further explore the driving mechanisms of F1’s motion before the interaction, we conducted a time-slice map of the LOS magnetogram along slice S2 (illustrated by the cyan line in Figure \ref{fig:fig2}(b2)), and the outcome is displayed in Figure \ref{fig:fig2}(e). From Figure \ref{fig:fig2}(e), it is evident that before the interaction, the P1 of the F1 continuously shifts northwestwards with a speed of about 0.36 km $s^{-1}$. Employing the DAVE4VM method \citep{Schuck...2006ApJ...646.1358S}, we computed the horizontal velocities from HMI vector magnetic field data with a 12-minute cadence. They are overlaid Bz maps in Figures \ref{fig:fig2}(c1-c4), where the blue and red arrows represent the horizontal velocities of the positive and negative magnetic fields, respectively. The velocity maps clearly show that the direction of motion of P1 is from southeast to northwest. This direction is similar to that of S2 in the LOS magnetogram and the orientation of F1 as it approached F2.

\begin{figure}
\plotone{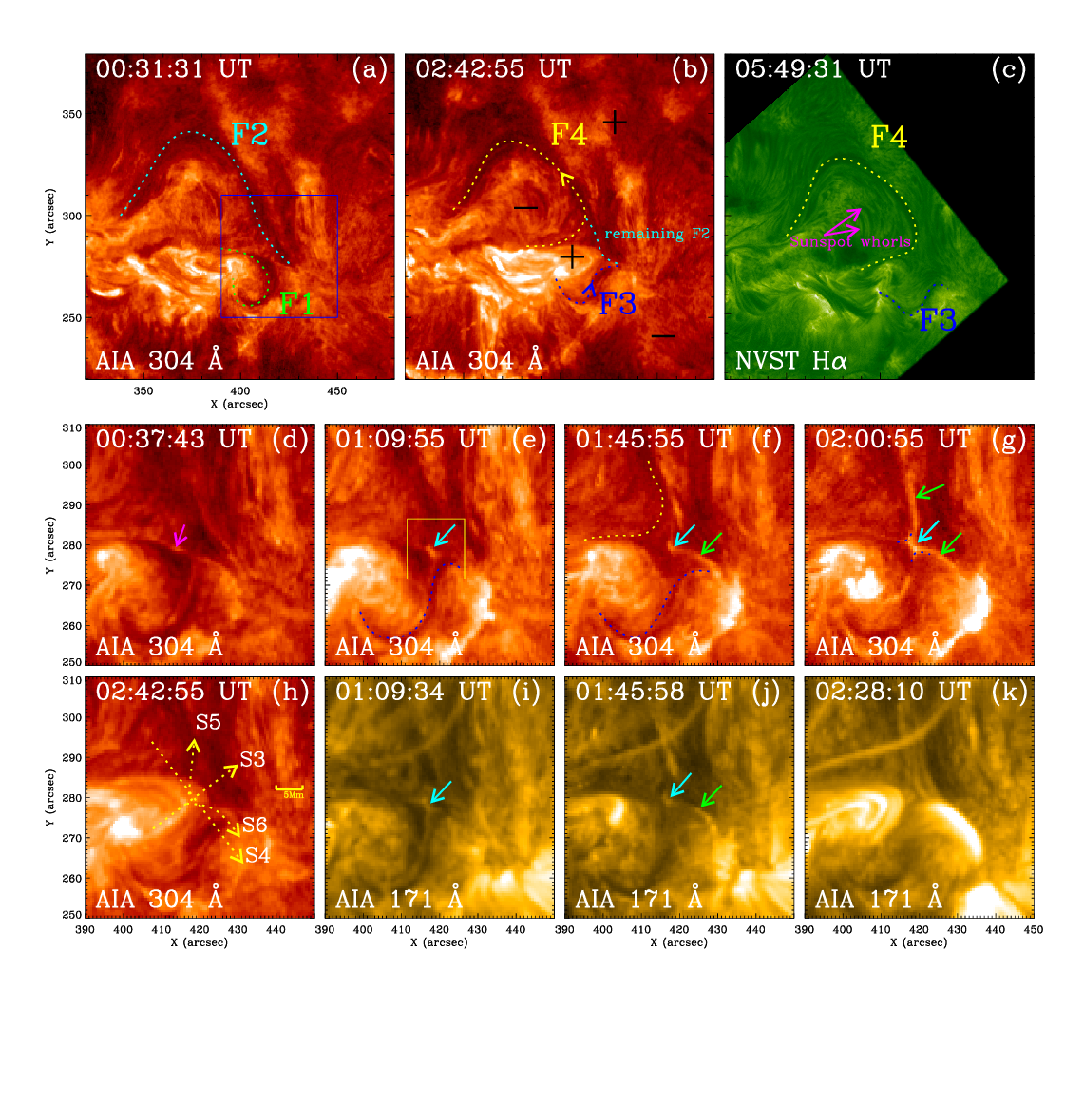}
\setlength{\abovecaptionskip}{-2.3cm}
\caption{The interaction process and magnetic reconnection features can be observed in the AIA 304 and 171 Å images (panels (a)–(b) and (d)–(k)). The NVST H$\alpha$ image illustrates the final morphology of F3 and F4 following the subsequent magnetic reconnection (panel (c)). The blue box in panel (a) indicates the views shown in panels (d)–(k) and Figure \ref{fig:fig7}. In panel (b), the directions of the axial field of F3 and F4 are represented by blue and yellow arrows, respectively.} PIasmoid is denoted by purple arrow in panel (d), while current sheets are represented by cyan arrows, and outflows are indicated by green arrows. The yellow box in panel (e) indicates the field of view of Figure \ref{fig:fig5}. Six time-slice maps obtained from the AIA 304 and 171 Å images along the four dotted lines in panel (h) are presented in Figure \ref{fig:fig4}. An animation of the 304 and 171 Å images shown in panels (a) and (d) is available. Its duration spans from 22:00 UT on May 8 to 06:00 UT on May 9. 
\label{fig:fig3}
\end{figure}

Figure \ref{fig:fig3} illustrates the interaction between the filaments F1 and F2 and the detailed process of magnetic reconnection. The interaction between the two filaments does not occur throughout the entire filaments, but rather in a small region (as shown in the blue box in Figure \ref{fig:fig3}(a)). Figures \ref{fig:fig3}(d)-(k) present the process of the interaction between F1 and F2 in detail. At 00:37 UT, a significant amount of plasma was observed to move along a part of filament F1 from the interaction region (denoted by the pink arrow in Figure \ref{fig:fig3}(d)). Subsequently, a brightening linear structure appeared in the interaction region around 01:09 UT (indicated by the cyan arrows in Figures \ref{fig:fig3}(e) and (i), and it may be the current sheet region. After 01:45 UT, a distinct current sheet structure was evident (As shown by the cyan arrows in Figures \ref{fig:fig3}(f) and (j)). At 02:00:55 UT, this structure became even more pronounced and larger, as seen in the 304 Å image (Figure \ref{fig:fig3}(g)). At both ends of the current sheet region, two bright cusp-shaped structures are visible (indicated by the blue dotted lines in Figures \ref{fig:fig3}(g). In the 304 and 171 Å images (Figures \ref{fig:fig3}(f), (g), and (j)), many hot outflows (green arrows) are observed intermittently. They exhibit elongated brightening plasma flows that gradually darken and disappear while moving away from the reconnection region. The outflows originate from the reconnection site and are more abundant and clearer towards P2 and N2, while it are rare and unclear towards P1 and N1. At 02:28 UT, the current sheet disappeared, but a significant brightening of the magnetic loops surrounding the interaction region was observed (as shown in Figure \ref{fig:fig3}(k)). The current sheet moved from the southeast to the northwest during the interaction, with a velocity of approximately 0.36 km $s^{-1}$. During the interaction, the period before 01:45 UT is defined as the weak reconnection phase, as the current sheet was unstable and small, while the period after 01:45 UT is defined as the main reconnection phase, as the current sheet became prominent and stable. A part of F1 connects with the right footpoint of F2 (as shown by the blue dashed lines in Figures \ref{fig:fig3}(e) and (f)), while another part of F1 connects with the left footpoint of F2 (as shown by the yellow dashed line in Figure \ref{fig:fig3}(f)), and the newly formed filaments are located away from the interaction region. These observational evidences suggest that magnetic reconnection occurs during the interaction of F1 and F2.

Ultimately, the footpoints of F1 and F2 are exchanged to form filaments F3 and F4, whose final outlines are shown in the 304 Å image (Figure \ref{fig:fig3}(b)). It is evident that the PILs associated with F3/F4 differ from those of F1/F2. In Figure 3(b), the directions of the axial fields for F3 and F4 are indicated by blue and yellow arrows, respectively. As viewed from the side of the positive polarity, the axial field directions of both F3 and F4 are directed to the left; thus, F3 and F4 are both sinistral filaments \citep{Martin...1994ASIC..433..303M}. In Figure 3(c), sunspot whorls rotating clockwise around the sunspot can be clearly observed. This suggests that the main spot retains positive magnetic helicity following the interaction \citep{Nakagawa...1971SoPh...19...72N,Rust.and.Martin...1994ASPC...68..337R,Chandra...2010SoPh..261..127C}. This supports the assertion that both F3 and F4 are sinistral filaments. After the interaction, F1 completely disappears. However, F2 is still partially connected by N2 and P2 (as indicated by the cyan dashed line in Figure \ref{fig:fig2}(b)), and this connection is maintained during 02:42 UT to 03:00 UT (as shown in the animation of Figure \ref{fig:fig2}), which indicates that F2 is not completely reconnected during the interaction. It is inferred that this reconnection represents a partial slingshot reconnection between two filaments with identical chirality \citep{Jiang...2013ApJ...764...68J}. The remaining part of F2 subsequently underwent reconnection with the magnetic loops connecting polarities N1 and P1 (see the following section), ultimately forming the final configurations of F3 and F4 (Figure \ref{fig:fig3}(c)). 

\begin{figure}
\plotone{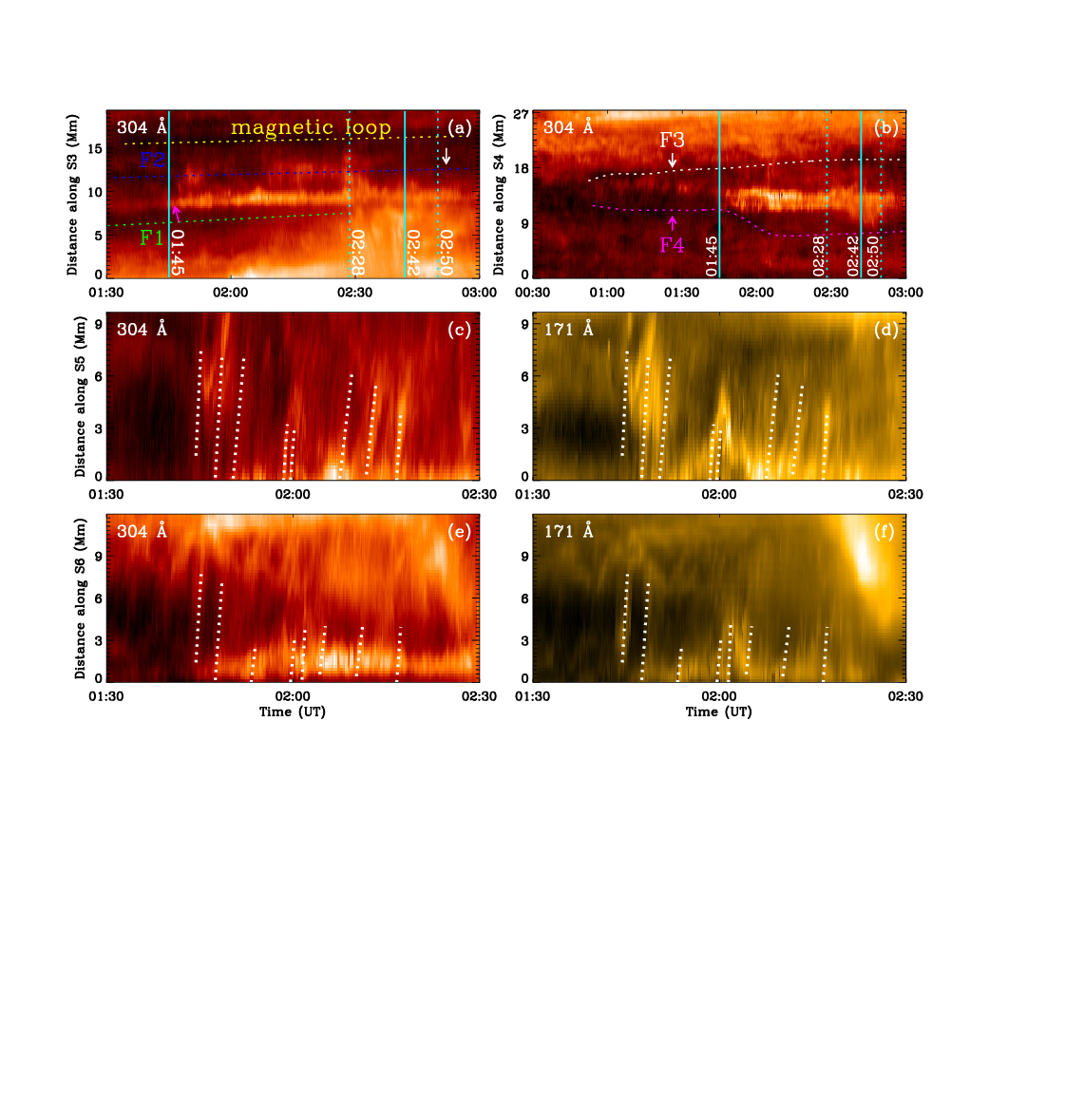}
\setlength{\abovecaptionskip}{-5.5cm}
\caption{(a)-(b) The time-distance diagrams along S5 and S6 in Figure \ref{fig:fig3}(h) employ AIA 304 Å images. The solid cyan lines represent the times at which significant stable current sheets appear and the formation times of F3 and F4, respectively. The dashed cyan lines denote the times at which the current sheets disappear and when the brightening of the loops ceases, respectively. (c)-(f) The time-distance diagrams along S3 and S4 in Figure \ref{fig:fig3}(h) utilize AIA 304 and AIA 171 Å images, respectively. The outflows are indicated by white dotted lines. 
\label{fig:fig4}}
\end{figure}

To investigate the details of two interactive filaments approach each other for magnetic reconnection, a time-slice map of the 304 Å images along slice S3 (as indicated by the yellow line in Figure \ref{fig:fig3}(h)) was conducted, and is displayed in Figure \ref{fig:fig4}(a). The motion trajectories of F1 and F2 during the reconnection are represented by the green and blue dashed lines, respectively.  It is shown that F1 and F2 are observed to be close to each other before 01:45 UT. A significant brightening is produced in the contact region of the two filaments at 01:45 UT (as indicated by the purple arrow in Figure \ref{fig:fig4}(a)), continuing until the current sheet disappears at 02:28 UT (Figure \ref{fig:fig3}(k)). It can be observed that after 02:42 UT, the region where F2 is located in Figure \ref{fig:fig4}(a) still shows a dark structure, supporting the conclusion that a part of F2 still exists even after the interaction has ended. To investigate the details of two newly formed filaments moving away during the interaction process, a time-slice map of the 304 Å images along slice S4 (as indicated by the yellow lines in Figure \ref{fig:fig3}(h)) was conducted, with the results displayed in Figure \ref{fig:fig4}(b). The white and the purple dotted lines denote the movement of F3 and F4, respectively. At about 00:50 UT, the two filaments begin to move apart, and from 01:45 UT, the two filaments are observed to be clearly moving away from each other, accompanied by a distinct current sheet. After 02:42 UT, F3 and F4 tend to become stable.

To study the dynamic characteristics of the outflows, four time-slice diagrams (Figure \ref{fig:fig4}(c)-(f)) were obtained along the slices S5 and S6 (indicated by the yellow dashed lines starting from the reconnection site in Figure \ref{fig:fig3}(h)) using the AIA 304 and 171 Å images. Eight typical outflow features marked by the white dashed lines in each time-slice diagram are selected, and their velocities are calculated. The speed of the outflows from reconnection site along left part of F2 range from 50.7 km $s^{-1}$ to 130.4 km $s^{-1}$ with an average velocity of 85.5 km $s^{-1}$ calculated from the 304 Å images, which is similar to that from the 171 Å images (Figures \ref{fig:fig4}(c)-(d)). Using the same method, their velocities were calculated to range from 58.5 km $s^{-1}$ to 159.9 km $s^{-1}$, resulting in an average velocity of 96.2 km $s^{-1}$.

\begin{figure}
\plotone{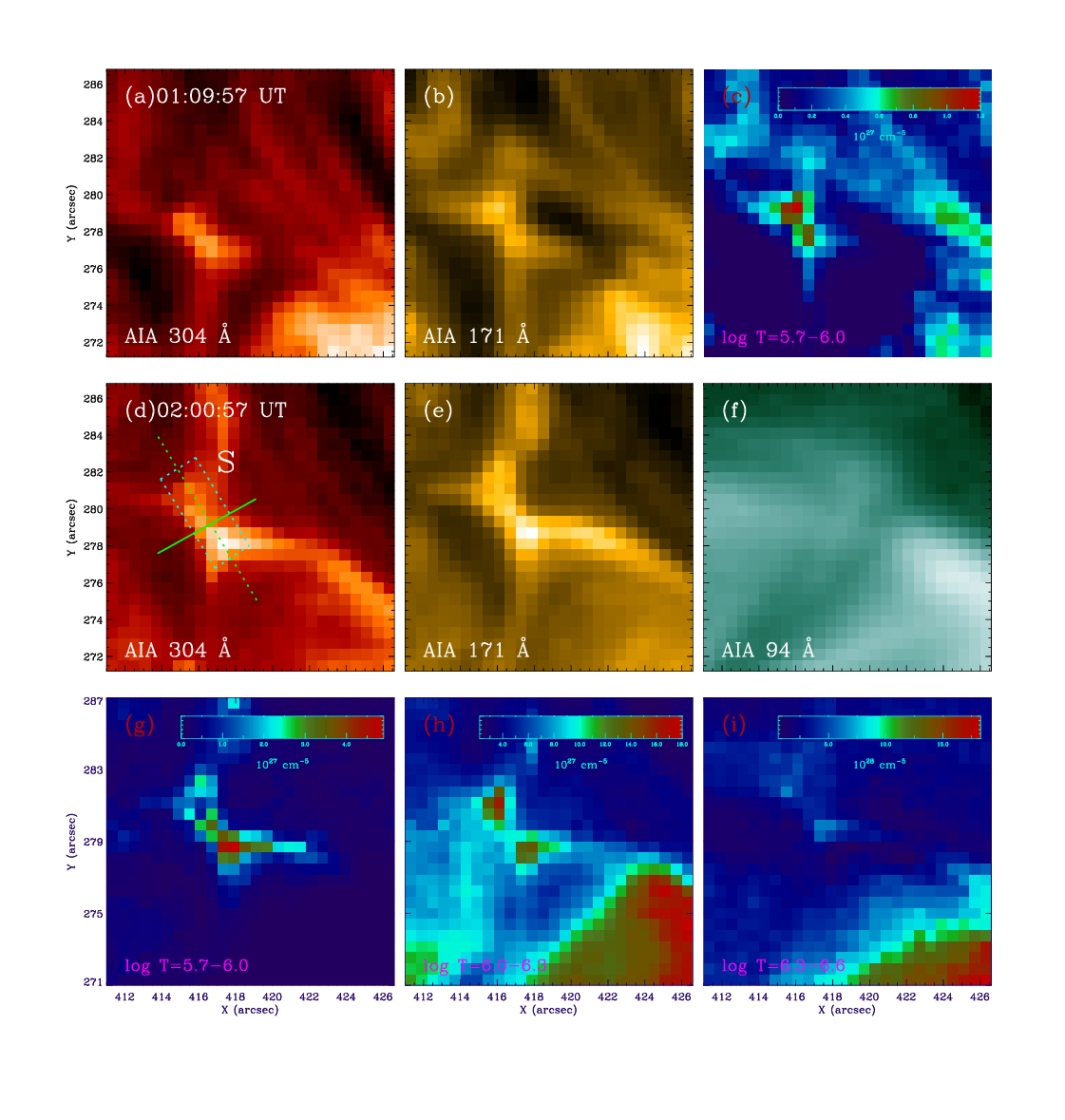}
\setlength{\abovecaptionskip}{-1.3cm}
\caption{Thermal characteristics of current sheets at 01:09 UT and 02:00 UT. (a)-(c) 304 Å  image, 171 Å  image, and emission measures at log T = 5.7-6.0 at 01:09:57 UT. (d)-(i) 304 Å  image, 171 Å  image, 94 Å  image, emission measures at log T = 5.7-6.0, emission measures at log T = 6.0-6.3, and emission measures at log T = 6.3-6.6 at 02:00:57 UT.
\label{fig:fig5}}
\end{figure}

Figure \ref{fig:fig5} illustrates the thermal characteristics of the current sheet. The typical current sheets at 01:09 UT during the phase of weak reconnection (Figures \ref{fig:fig5}(a)-(c)) and 02:00 UT during the main reconnection (Figures \ref{fig:fig5}(d)-(i)) were analyzed. Figures \ref{fig:fig5}(a)-(c) display the current sheet at 01:09:57 UT in the 304 Å image, the 171 Å image, and EMs at log T = 5.7-6.0, respectively. At this time, the current sheet exhibits strong emission in the range of log T = 5.7-6.0 (see Figure \ref{fig:fig5}(c)). However, the current sheet does not radiate strongly at higher temperatures. At 02:00 UT, the current sheet is observable in both the 304 Å and 171 Å images (Figures \ref{fig:fig5}(d) and (e)). However, it is not detectable in the 94 Å image (Figures \ref{fig:fig5}(f)). We extracted the intensity values along the green solid and dashed lines in Figure \ref{fig:fig5}(d) to calculate the length and width of the current sheet at 02:00:53 UT which are 2.18 Mm and 0.84 Mm, respectively. At this time, the entire current sheet demonstrates strong emission at log T = 6.0-6.3 (see Figure \ref{fig:fig5}(g)), while the ends of the current sheet show significant emission at log T = 5.7-6.0 (see Figure \ref{fig:fig5}(h)), and almost no emission is observed at log T = 6.3-6.6 (see Figure \ref{fig:fig5}(i)). The average EM of in the current sheet region marked by the box S in Figure \ref{fig:fig5}(d) is 1.21 $\times 10^{29}$ cm$^{-5}$. All the results suggest that the current sheet generated during the interaction between F1 and F2 is likely a low-temperature structure. The electron number density (\( n_e \)) of the current sheet can be estimated to be \( 3.79 \times 10^{10} \) cm\(^{-3}\) from the formula: $n_e = \sqrt{\frac{EM}{l}}
$ if assuming that the depth (l) of the current sheet equals its width \citep{Li_leping2021ApJ...908..213L}. This value is larger than the estimate of the electron number density for the current sheet provided by \cite{Li_leping2021ApJ...908..213L}.

\begin{figure}
\plotone{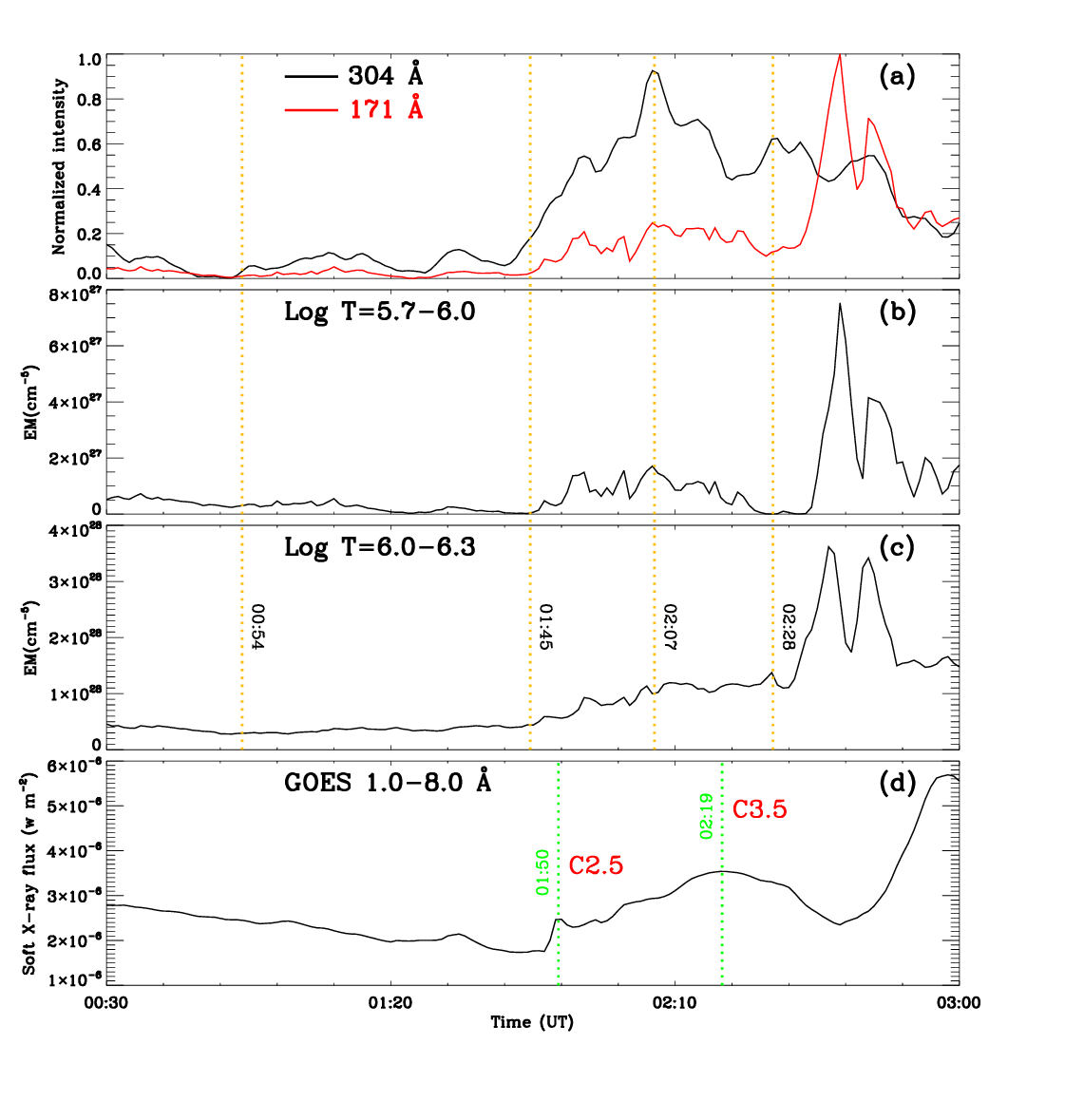}
\setlength{\abovecaptionskip}{-1.1cm}
\caption{Evolution of different parameters in the current sheets. (a)-(c) The normalized intensity and emission measures with time were calculated in the cyan dashed box S in Figure \ref{fig:fig5}(d). The yellow vertical dashed lines indicate the times corresponding to the onset of weak reconnection features, the onset of significant reconnection features, the moment when the current sheets are most pronounced, and the time of current sheet disappearance. (d) The profile of the GOES soft X-ray (SXR) flux in the 1.0-8.0 Å shows two C-class flares produced during the filaments interaction.}
\label{fig:fig6}
\end{figure}

To show the temporal evolution and physical parameters of the current sheet during reconnection, the normalized intensity (Figure \ref{fig:fig6}(a)) and EMs (Figures \ref{fig:fig6}(b) and (c)) with time were calculated in the current sheet region (marked by the cyan dashed box S in Figure \ref{fig:fig5})(d), respectively. They show that, at 00:54 UT, the EM values at log T = 5.7-6.0 and the intensity at 304 and 171 Å channel began to show a weak enhancement, and reached a minor peak at around 01:09 UT during weak magnetic reconnection. However, the EM at log T = 6.0-6.3 shows no significant variation. At around 01:45 UT, the intensity values and EM at log T =5.7-6.0 began to increase rapidly, may be due to the appearance of a stable and distinct current sheet structure in the magnetic reconnection region. They reach their peak at around 02:07 UT during the main magnetic reconnection period before starting to decline. From about 02:28 UT, the EM values at log T= 5.7-6.0 and intensity at 171 Å enhanced rapidly again, potentially due to adjacent heated structure approached the calculated area (as shown in Figure \ref{fig:fig3}(k)). Different from the Ems at log T = 5.7-6.0, the EMs at log T = 6.0-6.3 exhibited a slow enhancement prior to 02:28 UT, followed by a rapid increase. Two GOES class flares (C 2.5-class, started at 01:44 UT, peaked at 01:50 UT, ended at 02:10 UT; C 3.5-class, started at 01:52 UT, peaked at 02:19 UT, ended at 02:43 UT.) were detected during the reconnection, and their GOES SXR flux variations are shown in Figure \ref{fig:fig6}(d) (see the black curve). The peak emission of the flare (C 3.5) occurred after the peak intensity of the current sheet. Although the location of the flare (N13W25) was also within active region 13296, the reconnection site involved in the interaction of filaments was relatively distant, suggesting that the magnetic reconnection and the C 3.5 flare may not be related.

\begin{figure}
\plotone{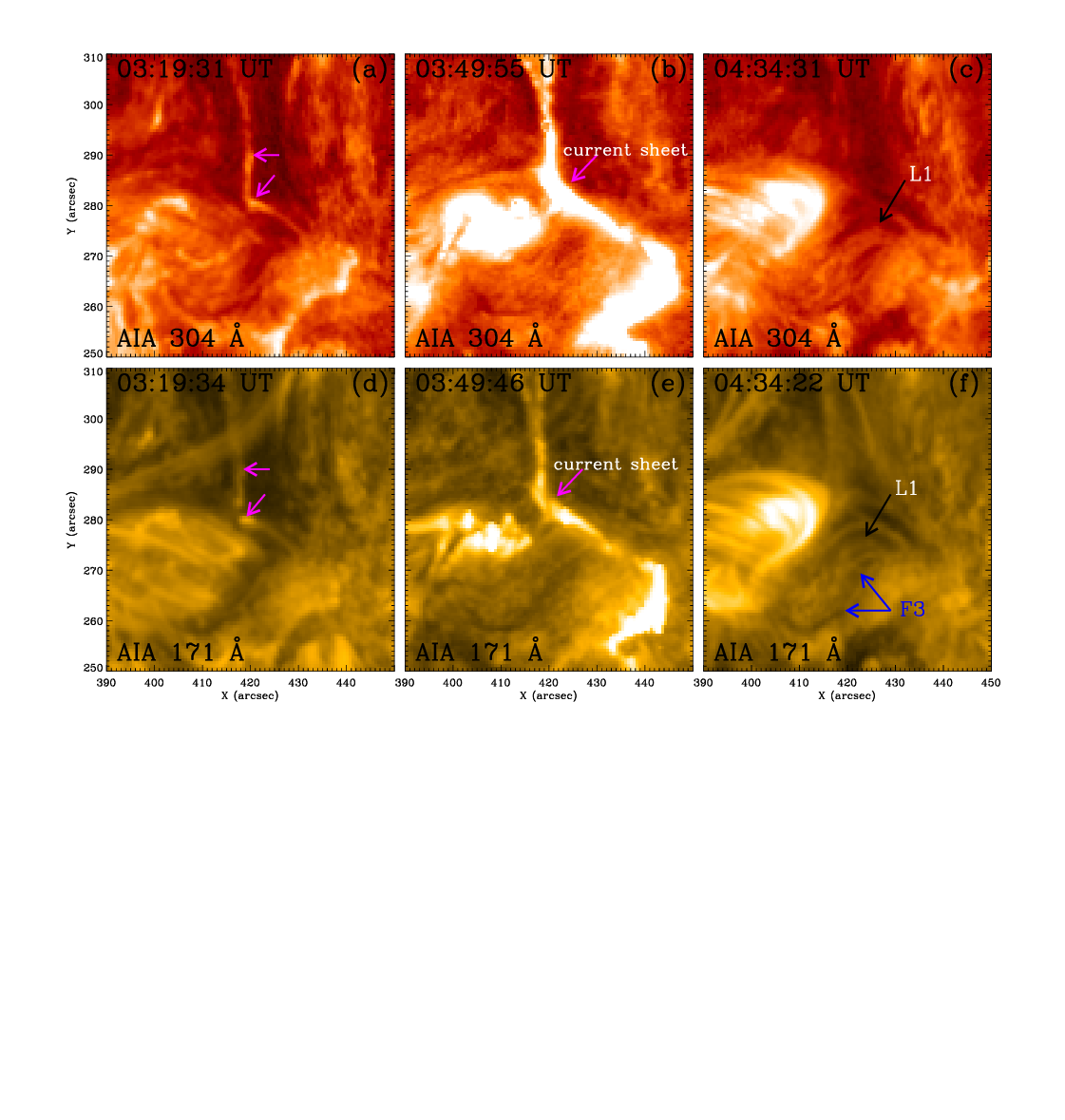}
\setlength{\abovecaptionskip}{-5.6cm}
\caption{Another magnetic reconnection in the current sheet region occurs after the interaction between F1 and F2, as observed in the AIA 304 Å and 171 Å images. Panels (a) and (d) illustrate the cusp structure and the movement of plasma blobs at the onset of reconnection. The wider and brighter current sheet structure is indicated by pink arrows in panels (b) and (e). The newly formed loop L1 following the magnetic reconnection is shown in panels (c) and (f).
\label{fig:fig7}}
\end{figure}

 Interestingly, after about 03:15 UT, a current sheet and plasma blobs are observed again in the same current sheet region (Figures \ref{fig:fig7}(a) and (d)), suggesting that another magnetic reconnection has occurred between the remaining portion of F2 and the magnetic loops connecting polarities N1 and P1. Subsequently, it was observed that the current sheet became wider and brighter (indicated by pink arrow in Figures \ref{fig:fig7}(b) and (e)). During this reconnection, a distinct magnetic loop L1 is newly formed at the northeast of F3 and does not merge with F3 (Figures \ref{fig:fig7}(c) and (f)). Another newly formed loops at the northwest of the reconnection site is merged with F4 and the material produced by the magnetic reconnection is continuously injected into F4 to form the final structure of F4 (Figure \ref{fig:fig3}(c)).

\section{CONCLUSIONS AND DISSCUSSIONS}\label{sec:conclusions and discussions}
We investigate the interaction between two active region filaments F1 and F2 and their successive magnetic reconnection processes in the AR NOAA 13296 and AR NOAA 13293 on May 9, 2023 utilizing the high spatial and temporal resolution observational data from SDO, NVST, and CHASE. The interaction is driven the motion of P1 with a velecity of 0.36 km $s^{-1}$. When F1 collides with F2, a magnetic reconnection occurs, producing current sheets with lower temperature at the junction of these two filaments. The current sheet continually slides from the southeast to the northwest as the northwest movement of the F1. It is measured to be 2.17 Mm in length and 0.84 Mm in width. Material outflows from two ends of the current sheet along the newly formed filament direction are observed, with velocity ranges of approximately 50.7 km $s^{-1}$ to 159.9 km $s^{-1}$. The interaction changes the magnetic connectivity of F1 and F2's magnetic field lines, eventually leading to the formation of two new filaments F3 and F4. After the interaction between F1 and F2, another magnetic reconnection occurred in the same current sheet region, possibly involving the reconnection of the magnetic loops connecting polarities N1 and P1 with the residual magnetic field of F2. 

It is worth exploring the mechanism that drives the interaction of filaments. Previous observations indicate two mechanisms that drive filament interactions: slow photospheric motions \citep{Kumar...2010ApJ...710.1195K,Chandra...2011SoPh..269...83C}, and the eruption of one interacting filament \citep{Jiang...2013ApJ...764...68J,Yang...2017ApJ...838..131Y}. P1 of one footpoint of F1 shifts from the southeast to the northwest at a speed of 0.36 km $s^{-1}$, while the filament F1 also moves from the southeast to the northwest at a speed of about 0.31 km $s^{-1}$. They are the same direction and are almost the same speeds. It implies that the motion of the P1 may have caused the motion of the F1, which is due to the evolution of the photospheric layer. The magnetic field in the photosphere acts on the P1 at one footpoint of F1 and also the filament F1, ultimately leading to the movement of filament F1 and its interaction with F2. Thus, this event is similar to the first mechanism; however, it does not cause large-angle rotation of the filament or a flare due to the interaction, which differs from the event described by \cite{Kumar...2010ApJ...710.1195K}. F2 remained nearly stationary; therefore, the merging of the two filaments is primarily due to the drag of P1 on F1, which ultimately results in a collision with F2 and leads to reconnection. During the interaction, the overall motion of the current sheet was calculated to be about 0.36 km $s^{-1}$. The direction of motion and speed of the current sheet are similar to those of F1; therefore, its movement may be caused by the motion of F1.

As suggested by \cite{Linton...2006JGRA..11112S09L}, due to the unequal axial flux of two flux tubes with the same helicity and low twist, the collision between the flux tubes likely led to only partial slingshot reconnection: the low-flux tube reconnected only with the outer shell of the high-flux tube, leaving it suspended from the interconnected core flux of the high-flux tube. In our event, it is showed that F2 did not completely reconnect with F1, and a part of F2 remains. Thus, we propose that the slingshot reconnection also occurred in the first magnetic reconnection and it may be a partial slingshot reconnection owing to unequal magnetic flux of F1 and F2. Additionally, in the simulated cases, the slingshot interaction occurs at interaction angles between $\frac{\pi}{2}$ and $\frac{3}{2} \pi$ \citep{Linton...2001ApJ...553..905L}.  In our event, the interaction angle of the filaments is approximately $\frac{5}{4} \pi$, which satisfies the angular requirement for slingshot reconnection in the simulations. Furthermore, unlike previous observations of slingshot reconnection events \citep{Kumar...2010ApJ...710.1195K,Chandra...2011SoPh..269...83C,Filippov...2011ARep...55..541F,Jiang...2013ApJ...764...68J,Yang...2017ApJ...838..131Y}, in this event, the distinct reconnection features and current sheets were identified in the interaction region. It marks the first observation of current sheet structure in filaments slingshot reconnection events. The process of magnetic reconnection leading to the formation of two identical chirality filaments (F3 and F4) during the interaction between two identical chirality filaments (F1 and F2) in this event is similar to events observed by \cite{Jiang...2013ApJ...764...68J}. However, our event differs from the interaction between an erupting filament and a non-erupting filament observed by \cite{Jiang...2013ApJ...764...68J}, instead, we observed the interaction of two non-erupting filaments driven by magnetic field evolution. Additionally, after partial slingshot reconnection, another magnetic reconnection occurs between the remaining part of F2 and the magnetic loops connecting magnetic polarities N1 and P1, continuing until F2 completely disappears. But in the event described by \cite{Jiang...2013ApJ...764...68J}, the remaining part of the large filament did not undergo external magnetic reconnection with other magnetic structures. The material generated by the reconnection is injected into F4, altering its morphology. The newly formed magnetic loop L1 is located above F3, and L1 is not merged into F3, indicating that the footpoints of L1 and F3 may be positioned at the same magnetic polarities but in different locations. Based on previous studies, we speculate that reconnection between flux ropes with unequal magnetic flux is more common in the corona than reconnection between flux ropes with equal magnetic flux, as unequal magnetic flux is more prevalent.

The authors sincerely appreciate the SDO, CHASE, and NVST teams for providing useful data. This work is sponsored by the Strategic Priority Research Program of the Chinese Academy of Sciences, Grant No. XDB0560000, by the National Science Foundation of China (NSFC) under grant Nos. 12325303, 11973084, 12003064, 11973088, 12203097, and 12473059, by the Yunnan Science Foundation of China under Nos. 202201AT070194, and 202101AT070032, by Key Research and Development Project of Yunnan Province under No. 202003AD150019, by the Youth Innovation Promotion Association, CAS (No. 2019061), CAS "Light of West China" Program, by the Yunnan Key Laboratory of Solar Physics and Space Science under No. 202205AG070009, by Yunnan Provincial Science and Technology Department under the number 202305AH340002, by Yunnan Science Foundation for Distinguished Young Scholars under No. 202001AV070004, by Yunnan Fundamental Research Projects under Nos. 202301AT070347, and 202301AT070349, by Yunnan Provincial Department of Education Science Research Fund Project under No. 2025J0945, by Chuxiong Normal University Doctoral Research Initiation Fund Project under No. BSQD2420, and by Yunnan Revitalization Talent Support Program. 

\vspace{5mm}
\bibliography{MS_21}
\bibliographystyle{aasjournal}

\end{document}